# A Random Shuffle Method to Expand a Narrow Dataset and Overcome the Associated Challenges in a Clinical Study: A Heart Failure Cohort Example

*Lorenzo Fassina*[1]\*, *Alessandro Faragli*[2,3,4,5], *Francesco Paolo Lo Muzio*[6,7], *Sebastian Kelle*[2,3,4,5], *Carlo Campana*[8], *Burkert Pieske*[2,3,4,5], *Frank Edelmann*[3,4,5] and *Alessio Alogna*[3,4,5]\*

[1] Department of Electrical, Computer and Biomedical Engineering, University of Pavia, Pavia, Italy, [2] Department of Internal Medicine and Cardiology, Deutsches Herzzentrum Berlin, Berlin, Germany, [3] Department of Internal Medicine and Cardiology, Charité – Universitätsmedizin Berlin, Berlin, Germany, [4] Berlin Institute of Health (BIH), Berlin, Germany, [5] DZHK (German Centre for Cardiovascular Research), Partner Site Berlin, Berlin, Germany, [6] Department of Surgery, Dentistry, Paediatrics and Gynaecology, University of Verona, Verona, Italy, [7] Department of Medicine and Surgery, University of Parma, Parma, Italy, [8] Department of Cardiology, Sant'Anna Hospital, ASST-Lariana, Como, Italy

Heart failure (HF) affects at least 26 million people worldwide, so predicting adverse events in HF patients represents a major target of clinical data science. However, achieving large sample sizes sometimes represents a challenge due to difficulties in patient recruiting and long follow-up times, increasing the problem of missing data. To overcome the issue of a narrow dataset cardinality (in a clinical dataset, the cardinality is the number of patients in that dataset), population-enhancing algorithms are therefore crucial. The aim of this study was to design a random shuffle method to enhance the cardinality of an HF dataset while it is statistically legitimate, without the need of specific hypotheses and regression models. The cardinality enhancement was validated against an established random repeated-measures method with regard to the correctness in predicting clinical conditions and endpoints. In particular, machine learning and regression models were employed to highlight the benefits of the enhanced datasets. The proposed random shuffle method was able to enhance the HF dataset cardinality (711 patients before dataset preprocessing) circa 10 times and circa 21 times when followed by a random repeated-measures approach. We believe that the random shuffle method could be used in the cardiovascular field and in other data science problems when missing data and the narrow dataset cardinality represent an issue.

## INTRODUCTION

Heart failure (HF) affects at least 26 million people worldwide (1), so predicting adverse events in HF patients represents a major target of clinical data science. Common challenges in clinical studies and trials are as follows (2, 3): (i) troubles in finding patients fitting the eligibility criteria (e.g., rare disease); (ii) difficulties in the enrollment because of a poorly formulated informed consent;

(iii) data collection problems; (iv) time delays because of complicated study design or due to unpredictable events; and (v) financial demands of the clinical practice. All these issues could be the cause of missing data and datasets with narrow cardinality, which are relevant challenges in data science (in a clinical dataset, the cardinality is the number of patients in that dataset).

As a consequence, researchers need to produce novel hypotheses and methods to deal with these issues, which are particularly critical when the dataset is used to build risk models in the field of clinical cardiology. A successful effort to overcome the abovementioned issues is represented by the MAGGIC risk score, developed as a tool of risk stratification for both morbidity and mortality in HF patients (4, 5). To build MAGGIC, Pocock et al. (5) have combined 30 datasets to enlarge patients' cardinality, thereby reaching an astonishing amount of 39,372 patients, and handled the missing patients' values via multiple imputations using chained equations (6, 7). In detail, to deal with missing data, regression equations are defined; the missing values are initially replaced by randomly chosen observed values of each variable, and then the missing values are replaced by a random draw from the distribution defined by the regression equations, and at the end of the last iteration, the final value becomes the chosen imputed value. Hence, we can argue that a random procedure could be important to overcome not only the issue of missing data, but also, at the same time, the one of narrow dataset cardinality.

The conceptual challenge of missing data is dual: 1) missing patients (i.e., completely missing data but plausible patients, as discussed later) who cause a narrow cardinality of the dataset and 2) missing data in patients with a partial list of needed values. In the current work, we unify the vision of these two kinds of missing data, searching for them with a random method, our novel random shuffle method without the use of specific hypotheses and regression models: we only need the original data, and we randomly shuffle them while it is statistically legitimate. "Statistically legitimate" means that, to validate our random shuffle method, the new datasets with enhanced cardinality were compared to those enhanced via an established random repeated-measures method (8, 9).

Indeed, the aim of this work is not to obtain a risk score, but to introduce an innovative method to enlarge the dataset cardinality and boost up the statistical performance. Our random shuffle method can be applied in other research fields when both missing data and limited dataset are issues because of financial, experimental, or ethical limitations.

## DATA AND METHODS

### Original Dataset

The clinical dataset is composed of a total of 711 German, Austrian, and Italian patients suffering from HF in different stages, in hospital facility due to either an acute hospitalization or an ambulatory visit, released and followed up for a period of 6 months. Patients were enrolled in two distinct clinical studies: (i) the Aldo-DHF trial (10), a multicenter, randomized, placebo-controlled, double-blind, two-armed, parallel-group study that enrolled patients from 10 trial sites in Germany and Austria (data are available in the **Supplementary Materials**) and (ii) the STOP-SCO trial, a prospective, multicenter, observational study that enrolled patients from 10 hospitals in the Northern Italy (unpublished data, that are available in the **Supplementary Materials**). The protocol and amendments were approved by the institutional review board at each participating center, and the trials were conducted in accordance with the principles of the Declaration of Helsinki, Good Clinical Practice guidelines, and local and national regulations. Written informed consent was provided by all patients before any study-related procedures were performed.

The studied endpoints at 6 months were a composite endpoint (all-cause hospitalization plus all-cause mortality) and all-cause hospitalization.

The dataset is organized in rows (patients) and columns (clinical parameters or features). The features are of two types:

i) 13 binary features that show the presence (value = 1) or the absence (value = 0) of the following conditions: peripheral edema, composite endpoint, age >75 years, angiotensin receptor blockers intake, β-blockers intake, left ventricular ejection fraction at admission >50%, nt-proBNP >1,000 pg/mL, diabetes, chronic kidney disease with glomerular filtration rate <50 mL/min, heart rate at release ≥90 bpm, anemia (hemoglobin concentration <12 g/dL for women, <13 g/dL for men), all-cause hospitalization endpoint, more than 2 hospitalizations in the last year; and

ii) 6 numerical features: age, heart rate at release, body weight at release, systolic aortic pressure at release, diastolic aortic pressure at release, left ventricular ejection fraction at admission.

To preprocess the clinical dataset for the removal of patients with missing values, two exclusion criteria were sequentially set: 1) at least an endpoint lacking (composite endpoint, all-cause hospitalization endpoint) and 2) at least a feature lacking (other than endpoints).

After the preceding data cleaning, the 13 binary features were used as dummy variables (11) to group the patients into classes, where the number of classes could be, at maximum, $2^{13}$. In particular, a self-balancing (12) (also called height-balancing) was applied to the tree of the binary features obtaining a new sorting of the dataset. In summary, the ordered list of the first 13 columns is the i) list above.

Moreover, because an intraclass-intrafeature random shuffling is possible if and only if the class cardinality is >1, the monoexample classes (i.e., with a lone patient) were excluded.

After preprocessing, the dataset is composed of 385 patients grouped into 61 classes. Conceptually, each class represents a particular clinical condition; in other words, the class label delimits a dataset subset inside which the shuffling is legitimate and not tautologic [as we show below, in statistical manner, via the comparison to a MATLAB-implemented repeated-measures fitting followed by its "random" method (8, 9); MATLAB®, The MathWorks, Inc., Natick, MA].

In **Figure 1A**, for demonstration purposes, we show a simplified representation of the original dataset with four patients analyzed with 3 features and grouped into 2 classes.

In **Figure 2A**, for sake of example and comparison with enhancing methods (**Figures 2B,C**), we plot two original numerical features for two classes (e.g., the 1st and the 3rd of 61 classes). The following sections will describe how to obtain variants of the original dataset.

### Repeated-Measure Variant

In MATLAB® (Statistics and Machine Learning Toolbox™), there are already implemented functions as the "fitrm" (acronym for "fit repeated-measures model") with the associated "random" method permitting to generate new random response values given predictor values (8, 9).

In particular, in the fitrm function, the measurements (the 6 numerical features above listed) are the responses, and the class column (with the aforementioned 61 classes) is the predictor variable. The fitrm function produces a repeated-measures model onto which we can apply the random method to randomly generate new response values, that is, new numerical measurements for our 6 numerical features. We called this random generation as "repeated-measures" variant (**Figure 1B**), and we added it to the original dataset (**Figure 1A**) obtaining an enhanced dataset (**Figure 2B**).

Theoretically, it is possible to generate at will without outputting replicated values, but we have introduced a calculus checkpoint to delete eventually replicated patients in the enhanced dataset.

### Shuffle Variant

In MATLAB®, we have implemented an intraclass random exchange/shuffle of values inside each feature (i.e., each feature is independently shuffled in random and intraclass manner). We called this random exchange/shuffle as "shuffle" variant (**Figure 1C**), and we added it to the original dataset (**Figure 1A**) obtaining an enhanced dataset (**Figure 2C**).

It is likely to shuffle with outputting replicated patients (especially inside low-cardinality classes), so we have introduced a calculus checkpoint to delete replicated patients in the enhanced dataset.

### Hotelling $t^2$ Statistic

Hotelling $T^2$ distribution is a multivariate distribution proportional to the F distribution; in particular, it is a generalization of the Student $t$ distribution for multivariate purposes. Hotelling $t^2$ statistic is a generalization of Student $t$ statistic used in multivariate hypothesis testing (13, 14).

In our multivariate problem, we have 6 numerical features, and we would enhance the original dataset without generating a different population ($p > 0.05$). So, the original dataset gives the expected multivariate mean vector (EMMV), and against EMMV, we compare the repeated-measures enhancement vs. the shuffle enhancement at a significance level of 0.05.

In other words, for the same enhanced number of patients, we are validating the shuffle enhancement using the repeated-measures enhancement which is an already accepted method: the shuffle enhancement is validated if and only if the $p$-value is not significant (i.e., the enhanced shuffled population is the same as the original dataset or the enhanced repeated-measures one).

### Combined Approach

In a combined approach, an enhanced shuffled population was subjected to a repeated-measures processing.

### Stressing the Enhanced Datasets via Machine Learning and Regression

In our specific cardiology problem (HF), the main goals of having enhanced datasets by enlarging their cardinality, while it is legitimate, are a greater classification/prediction skill (e.g., to predict the patient's class of risk) and a greater regression skill (e.g., to estimate the likelihood of two endpoints: composite endpoint, all-cause hospitalization endpoint). In other words, we are trying to overcome the issues of missing data and datasets with narrow cardinality, which are typically due to financial, experimental, or ethical limitations without losing the statistical nature of the original dataset, boosting its statistical performance while legitimate ($p > 0.05$ in $t^2$-test).

To highlight the benefits of the enhanced datasets vs. the original one, we have compared their classification/prediction skill and regression skill.

In detail, to stress via machine learning, we have used all the 19 features (13 binary, 6 numerical) and the column with the class labels as the response column (the enhanced dataset had 61 classes as the original one). A 10-fold cross-validation was applied to calculate the accuracy (%) by the MATLAB® Classification Learner application (methods: fine tree, fine KNN, weighted KNN, linear SVM; all default settings were unchanged).

To stress via regression, we have used 17 features (11 binary, i.e., excluding the 2 endpoints; 6 numerical) and, as response column, a column containing a specific endpoint (composite endpoint or all-cause hospitalization endpoint). A 10-fold cross-validation was applied to calculate the root mean square error (RMSE) by the MATLAB® Regression Learner application (methods: fine tree, linear, linear SVM; all default settings were unchanged).

## RESULTS

### Hotelling $t^2$ Statistic

The two enhanced populations (repeated-measure, shuffle) were the same as the original one until 20× enlargement; that is, we arrived up to 7,700 patients (including the 385 original). Further enhancements were not legitimate ($p < 0.05$).

In a combined approach, the preceding 20× shuffled population was subjected to a 2× repeated-measures processing,

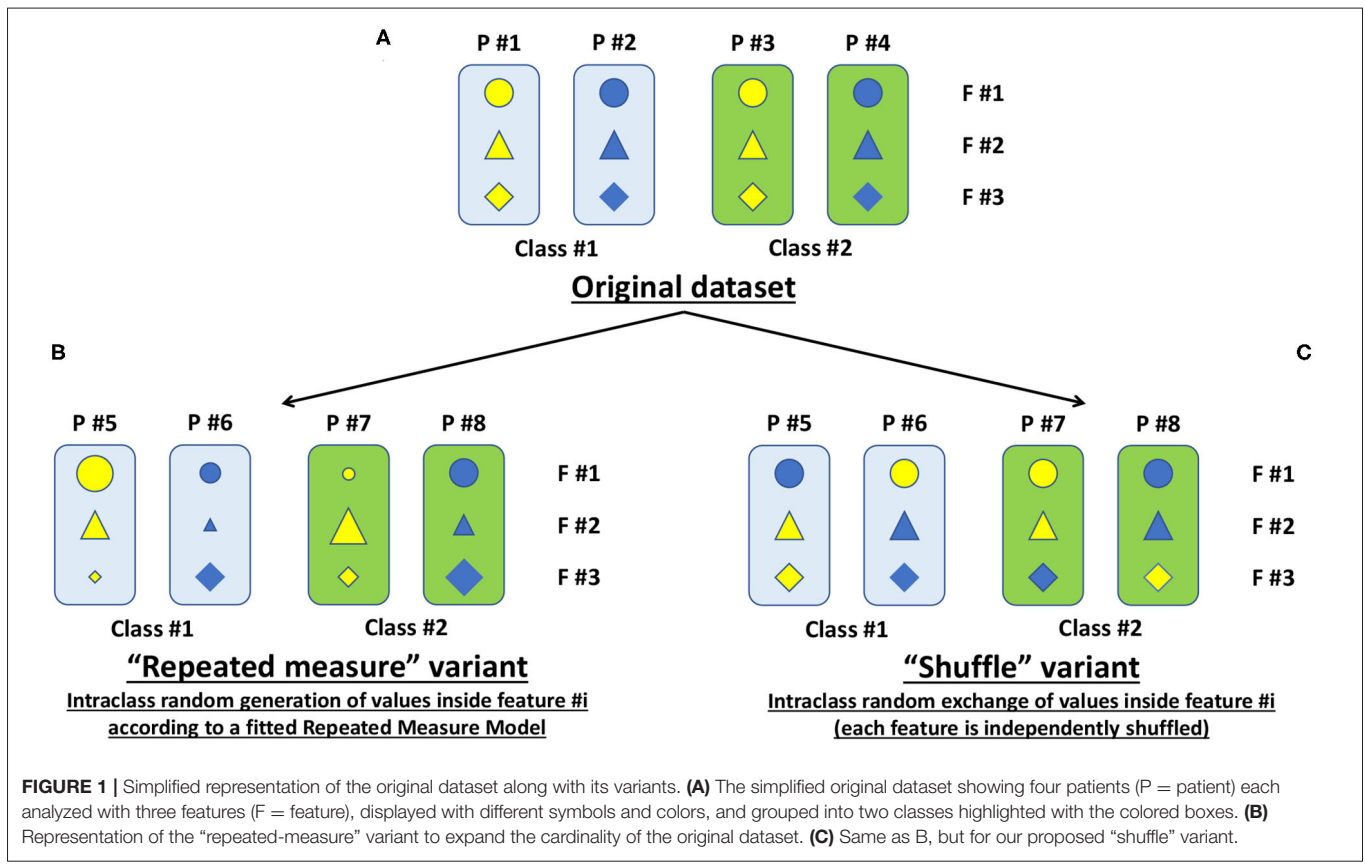

FIGURE 1 | Simplified representation of the original dataset along with its variants. (A) The simplified original dataset showing four patients (P = patient) each analyzed with three features (F = feature), displayed with different symbols and colors, and grouped into two classes highlighted with the colored boxes. (B) Representation of the "repeated-measure" variant to expand the cardinality of the original dataset. (C) Same as B, but for our proposed "shuffle" variant.

and we arrived up to 15,199 patients (including the 385 original). Further enlargements were not legitimate ($p < 0.05$).

## Stressing the Enhanced Datasets via Machine Learning and Regression

The comprehensive results are presented in the following tables in terms of accuracy (%) and RMSE.

Accuracy is a metric for evaluating the performance of machine learning in terms of the fraction of correct classifications. In this example dataset, high accuracy means that a sizable portion of patients was grouped into the correct classes (**Table 1**).

RMSE is a good estimator for the standard deviation of prediction errors; it informs about how far off we expect the regression model to be on its next prediction. If the RMSE is very small (**Tables 2**, **3**), the predicted value of an endpoint will practically coincide with the observed binary value in the future.

## DISCUSSION

To stratify patients according to their cardiovascular events risk in a 6-month follow-up after hospital discharge, the appropriate method of classification needs to be accurately determined in the case of the original dataset. In our case, the fine KNN algorithm implemented in MATLAB® revealed to be a good choice (accuracy equal to 93.2%, **Table 1**). However, the enlargement or enhancement of the cardinality of the original dataset, while it is legitimate, could possibly enable a greater classification/prediction skill. In detail, we have designed and developed a random shuffle method and validated it against the already used random repeated-measures method: the validation has given statistical legitimacy to the random shuffle method (while $p > 0.05$ via Hotelling $t^2$ statistic), and we have obtained a performance (accuracy up to 100%, independently from the classification method) better than that of the fine KNN dedicated only to the original dataset (**Table 1**). These results prove that the strategy with binary features, used to define the classes, and our random shuffle method to enhance the dataset can give a particularly good classification performance (**Table 1**).

To estimate the likelihood of the two endpoints (composite and all-cause hospitalization), a linear regression is already a good choice (**Tables 2**, **3**). However, the enlargement of the cardinality of the original dataset via both the random repeated-measures method and the random shuffle method or via the combined approach can give a better performance (RMSE down to 0), as stressed via the fine tree regression method. For example, a fatal clinical set is positive for nt-proBNP >1,000 pg/mL and heart rate ≥90 bpm, whereas a rehospitalization clinical set is positive for peripheral edema and left ventricular ejection fraction >50%, where the last parameter lightens the general health condition.

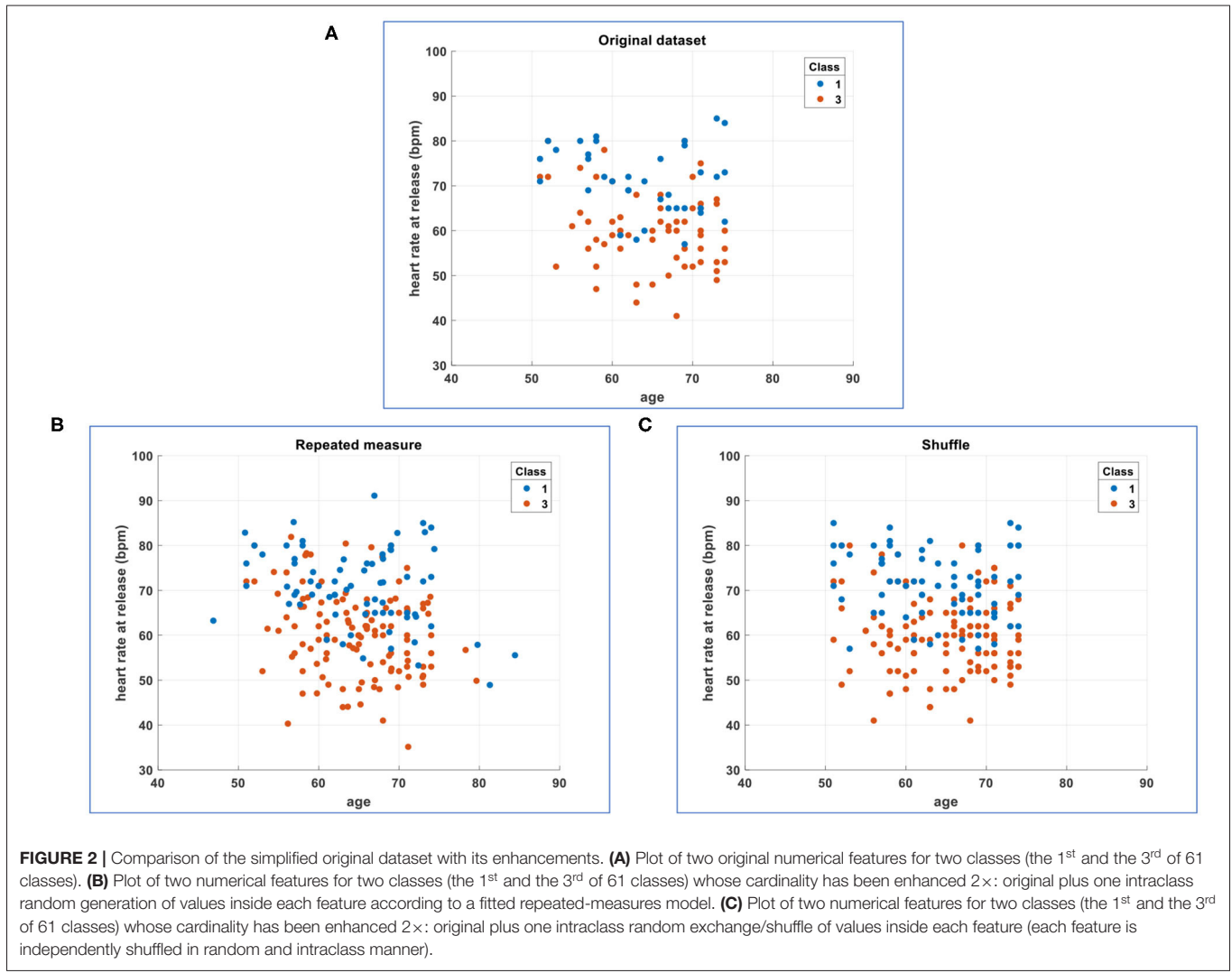

**FIGURE 2** | Comparison of the simplified original dataset with its enhancements. **(A)** Plot of two original numerical features for two classes (the 1st and the 3rd of 61 classes). **(B)** Plot of two numerical features for two classes (the 1st and the 3rd of 61 classes) whose cardinality has been enhanced 2×: original plus one intraclass random generation of values inside each feature according to a fitted repeated-measures model. **(C)** Plot of two numerical features for two classes (the 1st and the 3rd of 61 classes) whose cardinality has been enhanced 2×: original plus one intraclass random exchange/shuffle of values inside each feature (each feature is independently shuffled in random and intraclass manner).

**TABLE 1** | Machine learning with 10-fold cross-validation to calculate the classification accuracy (%).

| Accuracy (%) | 385 patients | 7,700 patients | 7,700 patients | 15,199 patients |
| --- | --- | --- | --- | --- |
|  | Original dataset | 20× Repeated measure | 20× Shuffle | Combined |
| Fine tree | 86.2 | 100 | 100 | 100 |
| Fine KNN | 93.2 | 100 | 100 | 100 |
| Weighted KNN | 86.0 | 100 | 100 | 100 |
| Linear SVM | 75.3 | 100 | 100 | 100 |

The names of the classification methods (fine tree, fine KNN, weighted KNN, linear SVM) refer to the preset tools inside the "Model Type" section of the MATLAB® Classification Learner application (all default settings were unchanged).

Clinicians could certainly claim that the abovementioned inferences could be easily made also without the use of mathematical methods or tools of artificial intelligence (e.g., classification/prediction or regression as shown in the **Tables 1–3**). Indeed, we consider such a provocative observation as a major strength of this study because we have validated the random shuffle method not only by statistics, but also, more importantly, by clinical judgment.

Another clinical strength is that the chosen features are patients' event ratios at hospitalization and follow-up. Thus, by randomly shuffling these features between patients, we are creating *in silico* plausible patients with a realistic and likely

**TABLE 2** | Regression with 10-fold cross-validation, endpoint = composite, to calculate the regression RMSE (root mean square error).

| RMSE | 385 patients | 7,700 patients | 7,700 patients | 15,199 patients |
| --- | --- | --- | --- | --- |
| | Original dataset | 20× Repeated measure | 20× Shuffle | Combined |
| Fine tree | 0.093 | 0 | 0 | 0 |
| Linear | $2.7 \times 10^{-16}$ | $3.2 \times 10^{-16}$ | $1.7 \times 10^{-15}$ | $2.5 \times 10^{-16}$ |
| Linear SVM | 0.108 | 0.066 | 0.065 | 0.065 |

The names of the regression methods (fine tree, linear, linear SVM) refer to the preset tools inside the "Model Type" section of the MATLAB® Regression Learner application (all default settings were unchanged).

**TABLE 3** | Regression with 10-fold cross-validation, endpoint = all-cause hospitalization, to calculate the regression RMSE (root mean square error).

| RMSE | 385 patients | 7,700 patients | 7,700 patients | 15,199 patients |
| --- | --- | --- | --- | --- |
| | Original dataset | 20× Repeated measure | 20× Shuffle | Combined |
| Fine tree | 0.003 | 0 | 0 | 0 |
| Linear | $1.9 \times 10^{-16}$ | $2.5 \times 10^{-16}$ | $6.8 \times 10^{-16}$ | $5.5 \times 10^{-16}$ |
| Linear SVM | 0.146 | 0.065 | 0.065 | 0.065 |

The names of the regression methods (fine tree, linear, linear SVM) refer to the preset tools inside the "Model Type" section of the MATLAB® Regression Learner application (all default settings were unchanged).

combination of comorbidities and event ratios. Therefore, the enhancement of the dataset cardinality yields not only statistical but also clinical worth.

In conclusion, we have shown that our random shuffle method is validated not only by statistical comparison to an already established method (the random repeated-measures method) but also, more notably, by the clinical knowledge and expertise. In addition, in comparison with the random repeated-measures method, a mathematical advantage of the random shuffle method is the absence of a fitting procedure. Consequently, we believe that our random shuffle method can also be applied in other research fields when missing data and the narrow cardinality of a dataset are issues because of financial, experimental, or ethical limitations.

## MORE TECHNICAL DISCUSSION

### Exclusion Criteria

Three exclusion criteria were sequentially set: 1) at least an endpoint lacking (thus, 116 patients were removed); 2) at least a feature lacking (other than endpoints) (another 67 patients removed); and 3) the monoexample classes (i.e., with a lone patient) were excluded (another 143 patients removed). Because the monoexample classes cannot be shuffled, one could certainly observe that exclusion criteria 1 and 2 are particularly selective. For instance, to increase the number of patients after preprocessing, only one endpoint at a time could be considered for patient's exclusion; this choice is certainly possible and correct, but implies the cutting of an entire feature, that is, the other endpoint, and as a consequence, we would obtain a reduced stratification of the patients. In addition, the random repeated-measures method does not tolerate missing data. Summarizing, the choice was (i) a lower number of patients but with all features, all endpoints, and full stratification or, on the contrary, (ii) a higher number of patients but with a reduced set of features and endpoints and with a reduced stratification. To stress the random shuffle method, we have chosen the first possibility, which is the "worst case" in terms of patients' number and stratification. In any case, the meaning of the random shuffle method remains the same as described above. Moreover, the choice permitted the use of the same data for both classification and regression.

### Cardinality Enhancement

The cardinality of the original dataset could be small because of two concomitant reasons: (i) a small number of classes (low stratification) and (ii) a small number of patients inside the classes. With these traits of the original database, the intraclass-intrafeature random shuffling has "suffocating borders" in which to act, and the database enhancement is also subjected to the deletion of repeated patients: in that case, we can hypothesize that the times of dataset enhancement is calmed down by the small cardinality of the original dataset. On the contrary, we see the maximum possibility of enhancement when the number of classes and the number of class patients are both high. On the other hand, we see intermediate possibilities when the classes are few but with many patients in each and, vice versa, when the classes are many but with few patients in each. In our original dataset, the classes were many (61 classes), and some of them had few patients (e.g., before cardinality enhancement, two or three or four patients); for additional details, see the following discussion dedicated to oversampling.

### Oversampling

The random shuffle method could also be seen as a new kind of oversampling dedicated to the classes of both minority (with low number of patients) and majority (with high number of patients). Oversampling is useful when there is an imbalance (related to the number of patients) between majority and minority classes able to downgrade the classification performance (15, 16). The imbalance can be corrected via oversampling inside minority classes and undersampling inside majority ones, e.g., via the SMOTE (Synthetic Minority Oversampling Technique) along with a randomly reduced number of patients in the majority classes (15). In a different approach respect to (15), where the information content is amplified or reduced in minority or majority classes, respectively, we have oversampled both minority and majority classes, while it is statistically legitimate; in other words, we preserve the imbalance (hallmark of a dataset), and we multiply the information content, while it is statistically legitimate, obtaining an enhanced classification and regression performance. We could also hypothesize that the reinforcement of all classes could improve the "exclusion power" of classification algorithms permitting them to better predict patients into reinforced minority classes.

## Cross-Validation for Oversampled Datasets

One could certainly observe that the cross-validation, although a very common and accepted technique to avoid the overfitting in classification and regression and so to ameliorate their prediction skill, could be prone to "overoptimism" when applied to oversampled datasets because similar samples or exact replicas may appear in both the training and test partitions. This issue has been clearly discussed by Santos et al. (17), who found a useful combination of characteristics to obtain a not-overoptimistic oversampling: (i) use of cleaning procedures, (ii) cluster-based synthetization of samples, and (iii) adaptive weighting of minority samples. The last cannot be applied because of the simple nature of the shuffling, but the other two have been comprised in the proposed method: the random shuffle is done in an intraclass manner, and then, we delete possible patients' replicas before further analysis; moreover, as third characteristic, each feature is independently shuffled, so that plausible patients are synthetized as clinically discussed above. The combination of these three method's traits makes us confident in the cross-validation done.

## CLINICAL LIMITATIONS

The clinical timepoint is to be considered approximately in the middle between those of the two trials used (Aldo-DHF and STOP-SCO). Even if the two trials were different in terms of patients' nationality, we used them together because they represent a real-life heterogeneous set of HF patients who are commonly observed in daily clinics. The risk prediction model at 6 months and an investigation on the differences between the data of the two trials were not purposes of this study and will be addressed in another work.

## DATA AVAILABILITY STATEMENT

Data and codes (with MIT License) along with reproducibility instructions are available in the **Supplementary Material** and also here: https://github.com/lorfas74/random-shuffle on GitHub development platform.

## ETHICS STATEMENT

The studies involving human participants were reviewed and approved by the institutional review board at each participating center and were conducted in accordance with the principles of the Declaration of Helsinki, Good Clinical Practice guidelines, and local and national regulations. The patients/participants provided their written informed consent to participate in this study.

## AUTHOR CONTRIBUTIONS

Random shuffle method (hypothesis, design, and implementation): LF. Statistics, machine learning, regression, and validation: LF, FPLM, AF, and AA. Acquisition of clinical data: AF, AA, SK, CC, FE, and BP. Clinical discussion: AF, AA, FPLM, FE, and BP. Wrote and edited the manuscript: all authors.

## FUNDING

AA is a participant in the BIH-Charité Clinician Scientist Program funded by the Charité – Universitätsmedizin Berlin and the Berlin Institute of Health. This work was supported by PRIN grant (2017AXL54F_002). We acknowledge support from the German Research Foundation (DFG) and the Open Access Publication Fund of Charité – Universitätsmedizin Berlin.

## SUPPLEMENTARY MATERIAL

The Supplementary Material for this article can be found online at: https://www.frontiersin.org/articles/10.3389/fcvm.2020.599923/full#supplementary-material